\documentclass[useAMS,usenatbib,usegraphicx]{mn2e}
\begin{document}
%
% Title
%
\title[Discovery of counter-rotating gas]{Discovery of
counter-rotating gas in the galaxies NGC~1596 and NGC~3203 and the
incidence of gas counter-rotation in S0 galaxies}
\author[M.\ Bureau and A.\ Chung]{M.\ Bureau$^{1}$\thanks{E-mail:
bureau@astro.ox.ac.uk} and A.\ Chung$^{2}$\thanks{E-mail:
archung@astro.columbia.edu}\\
$^{1}$Sub-Department of Astrophysics, University of Oxford, Denys
    Wilkinson Building, Keble Road, Oxford OX1~3RH, United Kingdom\\
$^{2}$Department of Astronomy, Columbia University, New York,
    NY~10027, U.S.A.}
\maketitle
%
% Abstract
%
\begin{abstract}
We have identified two new galaxies with gas counter-rotation
(NGC~1596 and NGC~3203) and have confirmed similar behaviour in
another one (NGC~128), this using results from separate studies of the
ionized-gas and stellar kinematics of a well-defined sample of $30$
edge-on disc galaxies. Gas counter-rotators thus represent $10\pm5\%$
of our sample, but the fraction climbs to $21\pm11\%$ when only
lenticular (S0) galaxies are considered and to $27\pm13\%$ for S0s
with detected ionized-gas only. Those fractions are consistent with
but slightly higher than previous studies. A compilation from
well-defined studies of S0s in the literature yield fractions of
$15\pm4\%$ and $23\pm5\%$, respectively. Although mainly based on
circumstantial evidence, we argue that the counter-rotating gas
originates primarily from minor mergers and tidally-induced transfer
of material from nearby objects. Assuming isotropic accretion, twice
those fractions of objects must have undergone similar processes,
underlining the importance of (minor) accretion for galaxy
evolution. Applications of gas counter-rotators to barred galaxy
dynamics are also discussed.
\end{abstract}
%
% Introduction
%
\section{INTRODUCTION}
\label{sec:intro}
It has been already almost $20$ years since the phenomenon of
counter-rotation in disc galaxies was discovered \citep{g87}, but both
the exact incidence and the origin of counter-rotating gas and stars
remain to be clarified. Most statistical studies indicate that roughly
$20$--$25\%$ of all lenticular (S0) galaxies with detected ionized-gas
(usually observed through optical emission lines) contain a
non-negligible fraction of gas-stars counter-rotation
\citep*[e.g.][]{bbz92,kfm96,kf01,pcvb04}, typically in the central
regions. The fraction is much lower in later type systems
\citep[e.g.][]{kf01,pcvb04}, presumably because co-rotating and
counter-rotating gas can not coexist at a given radius. Galaxies with
counter-rotating H$\,${\small I} or CO gas are also known (e.g.\
NGC~4826, \citealt*{bwk92}; NGC~3626, \citealt*{gsb98}), but to our
knowledge no sound statistics exists. The fraction of S0s with
stars-stars counter-rotation is also much lower (at most $10\%$ but
perhaps lower; \citealt{kfm96,pcvb04}), perhaps due to the increased
difficulty of detecting small numbers of counter-rotating
stars. Generally, however, the number of objects on which well-defined
studies are based is still rather small, and it is still the case that
most known counter-rotators were discovered fortuitously in targeted
studies (e.g.\ NGC~4546, \citealt{g87}; NGC~4138, \citealt*{jbh96}).

Our goal in this paper is thus to improve the (ionized-gas)
counter-rotation statistics in S0 galaxies and summarize the current
situation. Since counter-rotating material is a strong argument in
favour of hierarchical galaxy formation scenarios, this is an
important goal. In the interest of conciseness, we do not discuss here
the issue of counter-rotation in elliptical galaxies, although it is
probably related (see \citealt{r94} and \citealt{s98} for general
reviews of galaxies with misaligned angular momenta).

In Section~\ref{sec:obs}, we describe gas and stellar kinematics
observations of a large sample of edge-on disc galaxies. The
structure, kinematics, and environment of $2$ newly discovered
ionized-gas counter-rotators and $1$ known object are discussed in
detail in Section~\ref{sec:galaxies}. In Section~\ref{sec:discussion},
we quantify the incidence of counter-rotators using all available
observations and discuss their possible origin through continuous gas
infall and/or discrete gas and satellites accretion. Applications of
gas counter-rotators to barred galaxy dynamics are also discussed. We
conclude briefly in Section~\ref{sec:conclusions}.
%
% Observations
%
\section{OBSERVATIONS}
\label{sec:obs}
\subsection{Sample and data reduction}
\label{sec:sample}
Our sample consists of $30$ edge-on disc galaxies from the catalogs of
galaxies with a boxy or peanut-shaped (B/PS) bulge of \cite{j86},
\cite{s87} and \cite{sa87}, and from the catalog of galaxies with an
extreme axial ratio of \cite*{kkp93}. $80\%$ have a B/PS bulge and
roughly three-quarters either have probable companions or are part of
a group or cluster, although some are probably chance alignments so
the sample is not biased either against or for galaxies in a dense
environment. We warn that the usual morphological classification into
Hubble type \citep[e.g.][]{s61} is uncertain for edge-on galaxies,
since the bulge-to-disc ratio is effectively the only usable
criterion.

The observations from the Double Beam Spectrograph on the 2.3-m
telescope at Siding Spring Observatory were already discussed
elsewhere. The red arm of the spectrograph, used to study the
ionized-gas kinematics of the sample galaxies \citep{bf99}, was
centred on the H$\alpha$ $\lambda6563$ emission line, covering roughly
$950$~\AA\ with a spectral resolution of $1.1$~\AA\
($50$~km~s$^{-1}$). The blue arm, used for the stellar kinematics
\citep{cb04}, was centred on the Mg$b$ $\lambda5170$ absorption
triplet, covering again roughly $950$~\AA\ with a $1.1$~\AA\
($50$~km~s$^{-1}$) spectral resolution. The seeing was typically
$1.0$--$1.5$~arcsec. All data were reduced and analysed using standard
software and methods and we refer to \cite{bf99} and \cite{cb04} for
further details. We note that we have not extracted ionized-gas
rotation curves from our data, the entire two-dimensional spectra or
position-velocity diagrams (PVDs) constituting our final products.
%
% Figure: Kinematics
%
\begin{figure}
\begin{center}
\includegraphics[width=7.4cm]{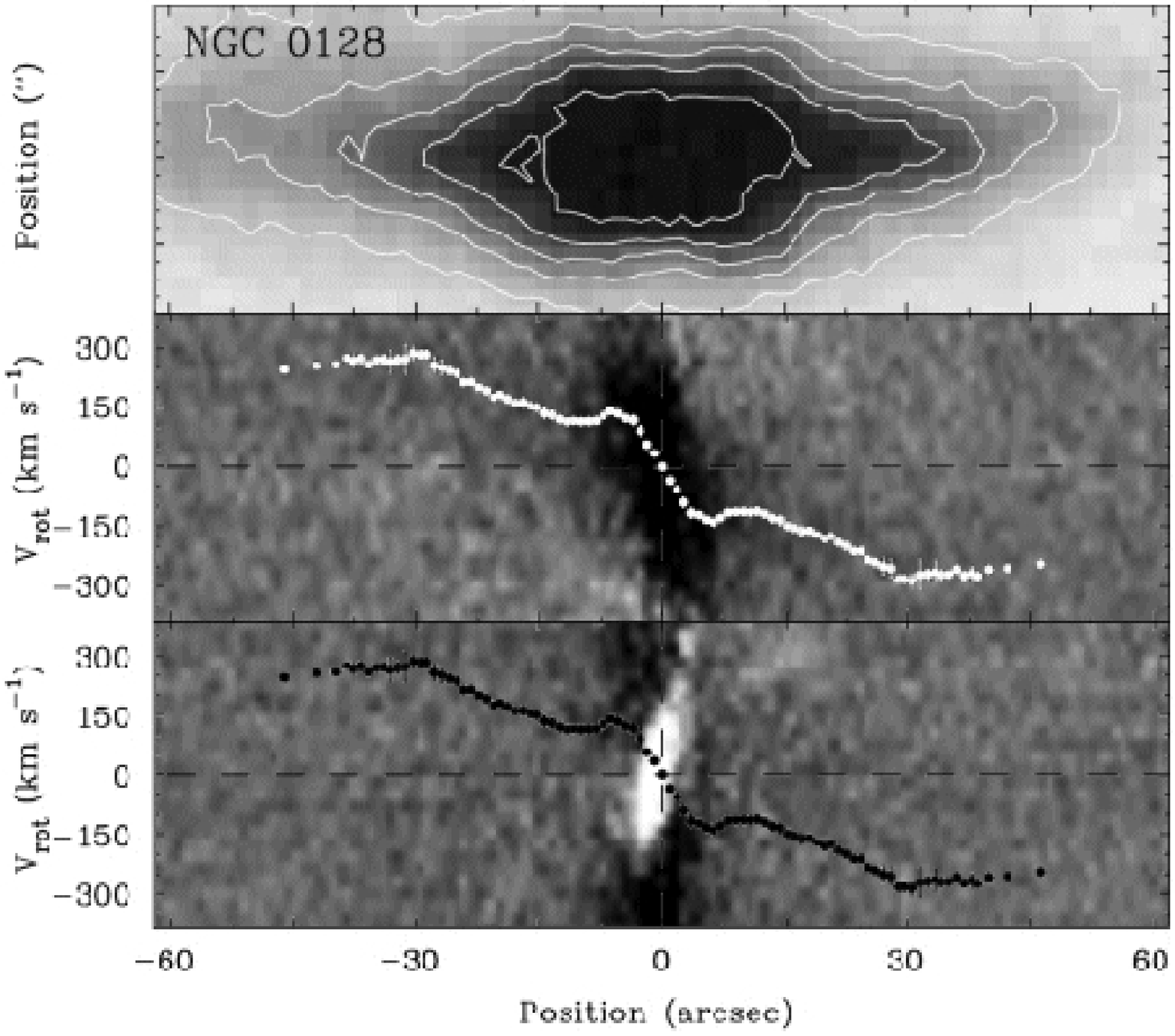}\\
\includegraphics[width=7.4cm]{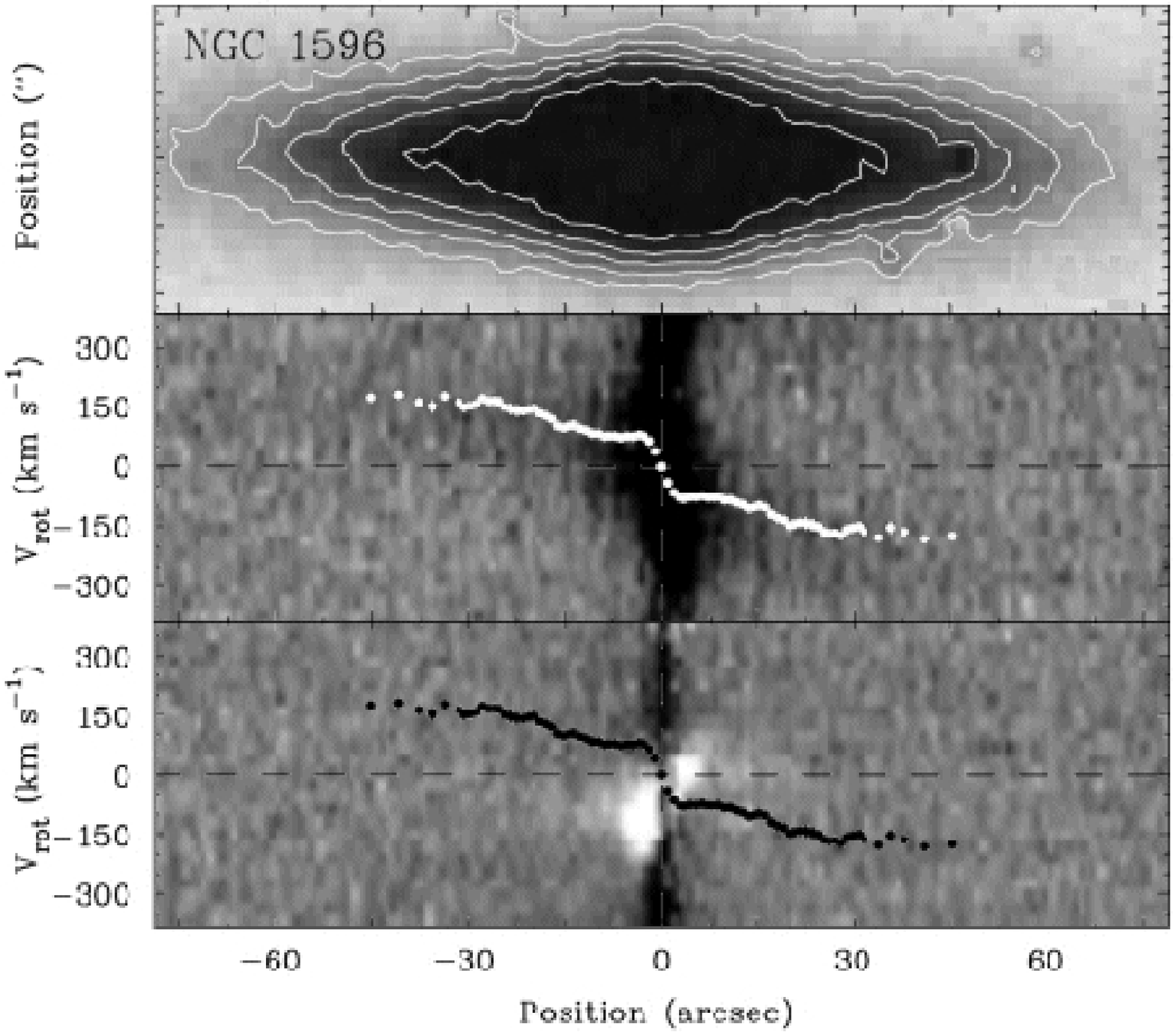}\\
\includegraphics[width=7.4cm]{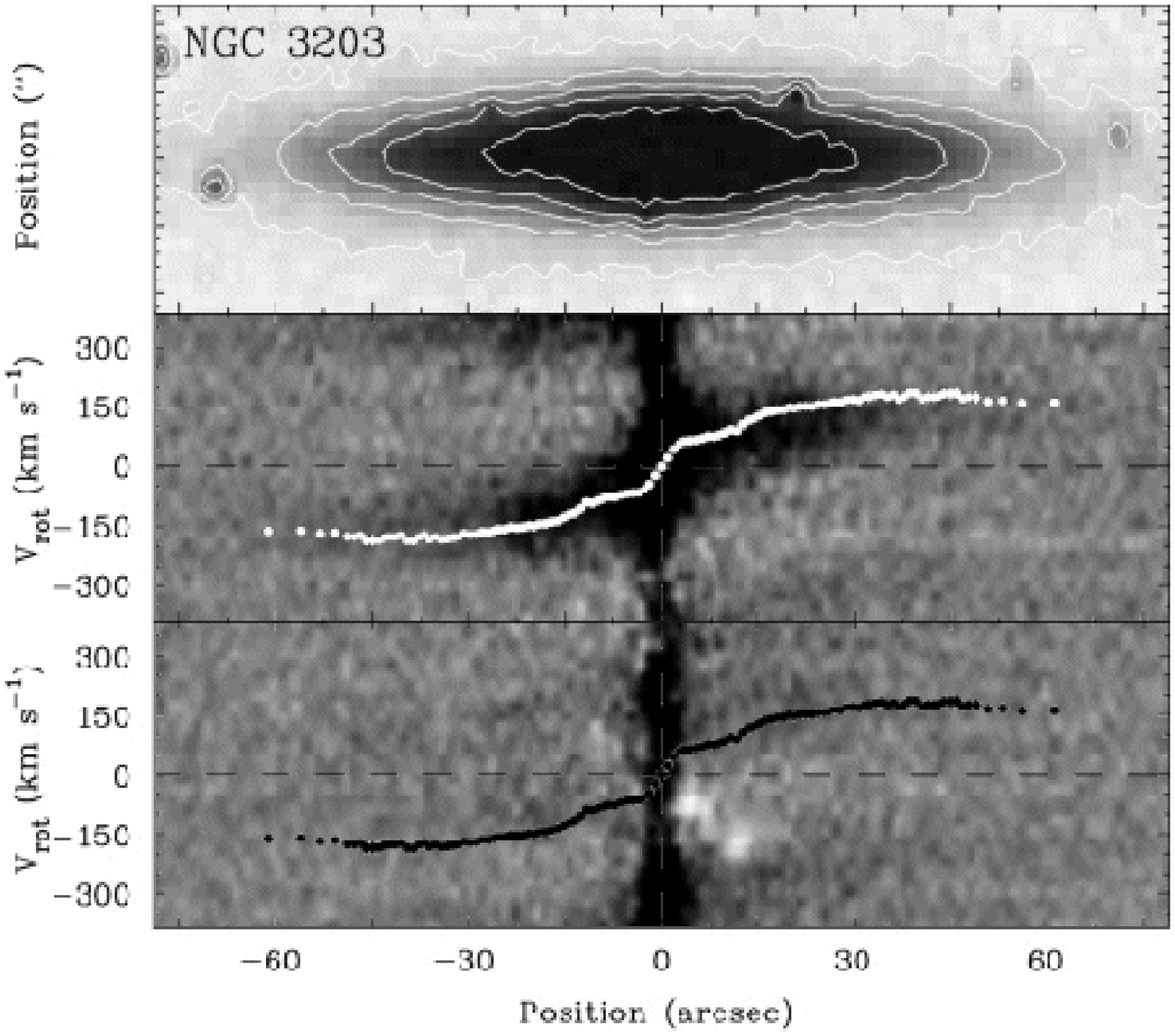}
\caption{Stellar and ionized-gas kinematics of NGC~128 (top),
NGC~1596 (middle) and NGC~3203 (bottom). Each panel shows, from top to
bottom and spatially registered, a grayscale image of the galaxy (with
contours) from the Digitized Sky Survey (DSS), the stellar rotation
curve (white dots) overplotted on a stellar absorption line, and the
stellar rotation curve (black dots) overplotted on an emission line
position-velocity diagram (here [OIII] $\lambda5007$). For all three
galaxies, the ionized-gas is clearly limited to the central regions
and counter-rotating with respect to the stars.}
\label{fig:kin}
\end{center}
\end{figure}
%
% Table: Basic Properties
%
\begin{table*}
\begin{minipage}{126mm}
\caption{Basic properties of the galaxies with gas counter-rotation}
\label{tab:prop}
\begin{tabular}{@{}lrrrr}
\hline
Quantity & NGC~128 & NGC~1596 & NGC~3203 & Source\\
\hline
Right ascension (J2000) & $00^{\rm h} 29^{\rm m} 15\fs0$ & $04^{\rm h} 27^{\rm m} 38\fs1$ & $10^{\rm h} 19^{\rm m} 33\fs8$ & NED \\
Declination (J2000) & $+02\degr 51\arcmin 51\arcsec$ & $-55\degr 01\arcmin 40\arcsec$ & $-26\degr 41\arcmin 56\arcsec$ & NED \\
Heliocentric velocity (km~s$^{-1}$) & $4241$ & $1510$ & $2424$ & NED \\
Distance (Mpc) & $60.3$ & $17.5$ & $32.4$ & HyperLeda \\
Morphological type & S0 pec & SA0: & SA(r)0$^+$? & NED \\
Apparent diameter (arcmin; $\mu_{B}=25$~mag~arcsec$^{-2}$) & $3.0\times0.9$ & $3.7\times1.0$ & $2.9\times0.6$ & NED \\
Total apparent $B$ magnitude & $12.7$ & $12.0$ & $13.0$ & HyperLeda \\
Total absolute corrected $B$ magnitude & $-21.4$ & $-19.3$ & $-19.9$ & HyperLeda \\
\hline
\end{tabular}
NED: NASA/IPAC Extragalactic Database: http://nedwww.ipac.caltech.edu/index.html\\
HyperLeda: http://www-obs.univ-lyon1.fr/hypercat/ 
\end{minipage}
\end{table*}
\subsection{Results}
\label{sec:results}
By comparing the ionized-gas and stellar kinematics, $3$ galaxies out
of the sample of $30$ show clear ionized-gas counter-rotation. That
is, the ionized-gas is rotating in the opposite direction to the bulk
of the stars. Those galaxies, along with their ionized-gas and stellar
kinematics, are shown in Figure~\ref{fig:kin}. We note that the
orientation of the slit was inverted for NGC~128 and NGC~1596 in
\cite{cb04}, and it has been corrected here. This has however no
bearing on the detection of counter-rotation. Basic properties of the
galaxies are listed in Table~\ref{tab:prop}. NGC~128 is a well-studied
case \citep[e.g.][]{ea97,ocmzb99}, but counter-rotation in NGC~1596
and NGC~3203 is reported here for the first time.

As in many previous cases \citep[e.g.][]{bbz92}, the counter-rotating
gas is limited to the inner bulge-dominated region of the galaxies,
with radial extent of order $1$~kpc. However, since the ionized-gas is
clearly asymmetric, since its extent clearly depends on the depth of
our spectra and on our ability to recover emission lines, and since
our spectra are not flux calibrated, we refrain from discussing the
exact sizes and masses of the counter-rotating discs.

Suffice it to say that while the ionized-gas appears fairly regular in
NGC~128 and NGC~1596, it is rather unsettled in NGC~3203. Except for a
hint of ionized-gas at large positive velocities in NGC~128, the
ionized-gas detected always appears confined to a region with rising,
approximately solid-body rotation. We refrain from plotting velocity
dispersion measurements, as the emission lines are generally
unresolved spectrally. We also note that our long-slit observations
can not constrain the exact alignment of the ionized-gas. The observed
counter-rotation only implies a kinematic (and angular momentum)
misalignment between the stars and ionized-gas between $90$ and
$270\degr$.

While the ionized-gas in NGC~128 is consistent with being in a ring
(which, if circular, yields a straight line in a PVD), that in
NGC~1596 clearly is not, or at least the ring must have a significant
width. A ring of ionized-gas in NGC~3203 could only explain the
observations if very patchy; it is more likely that the gas is
substantially disturbed.

\begin{figure}
\begin{center}
\includegraphics[width=6.9cm]{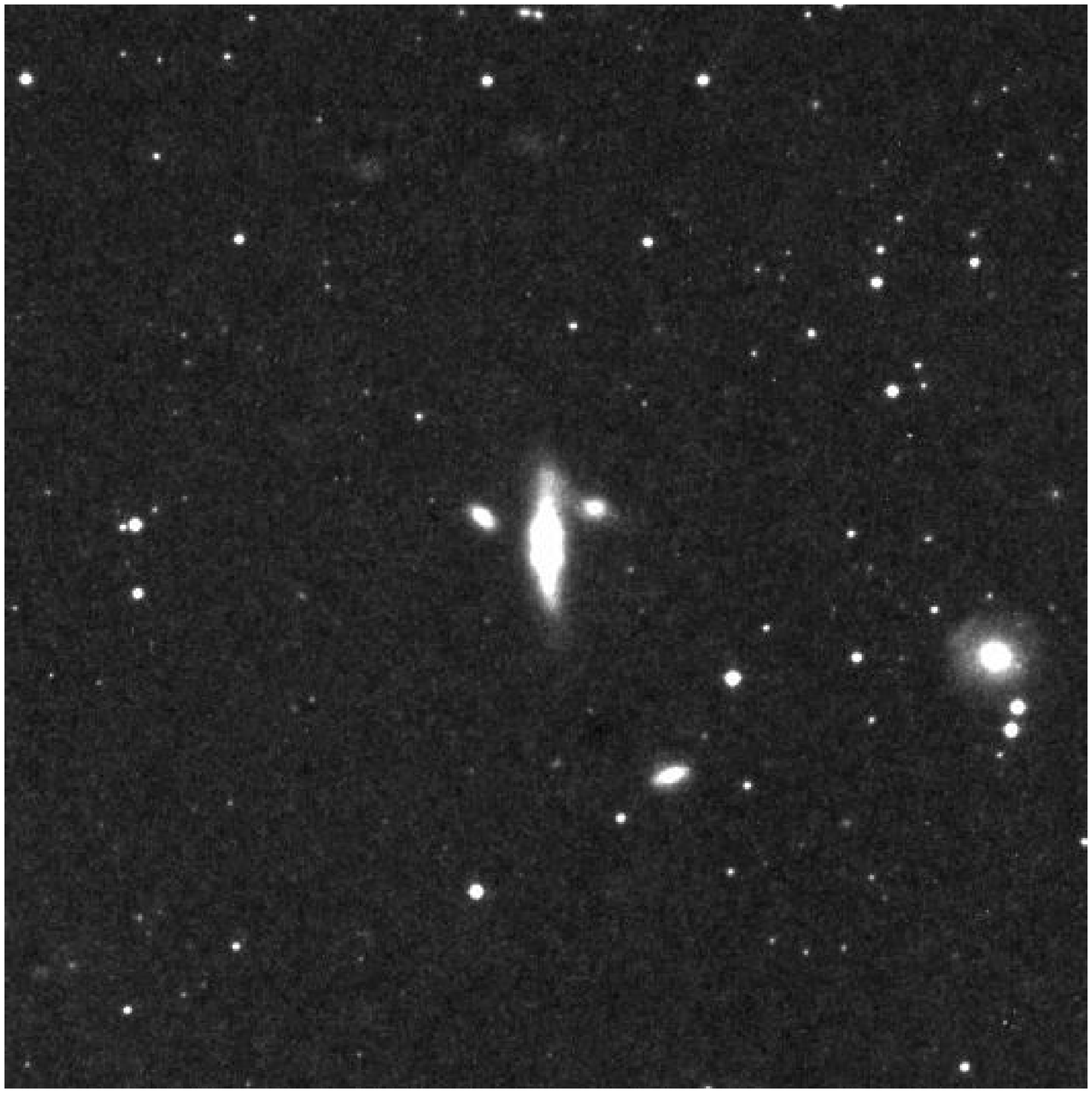}\\
\vspace*{0.5cm}
\includegraphics[width=6.9cm]{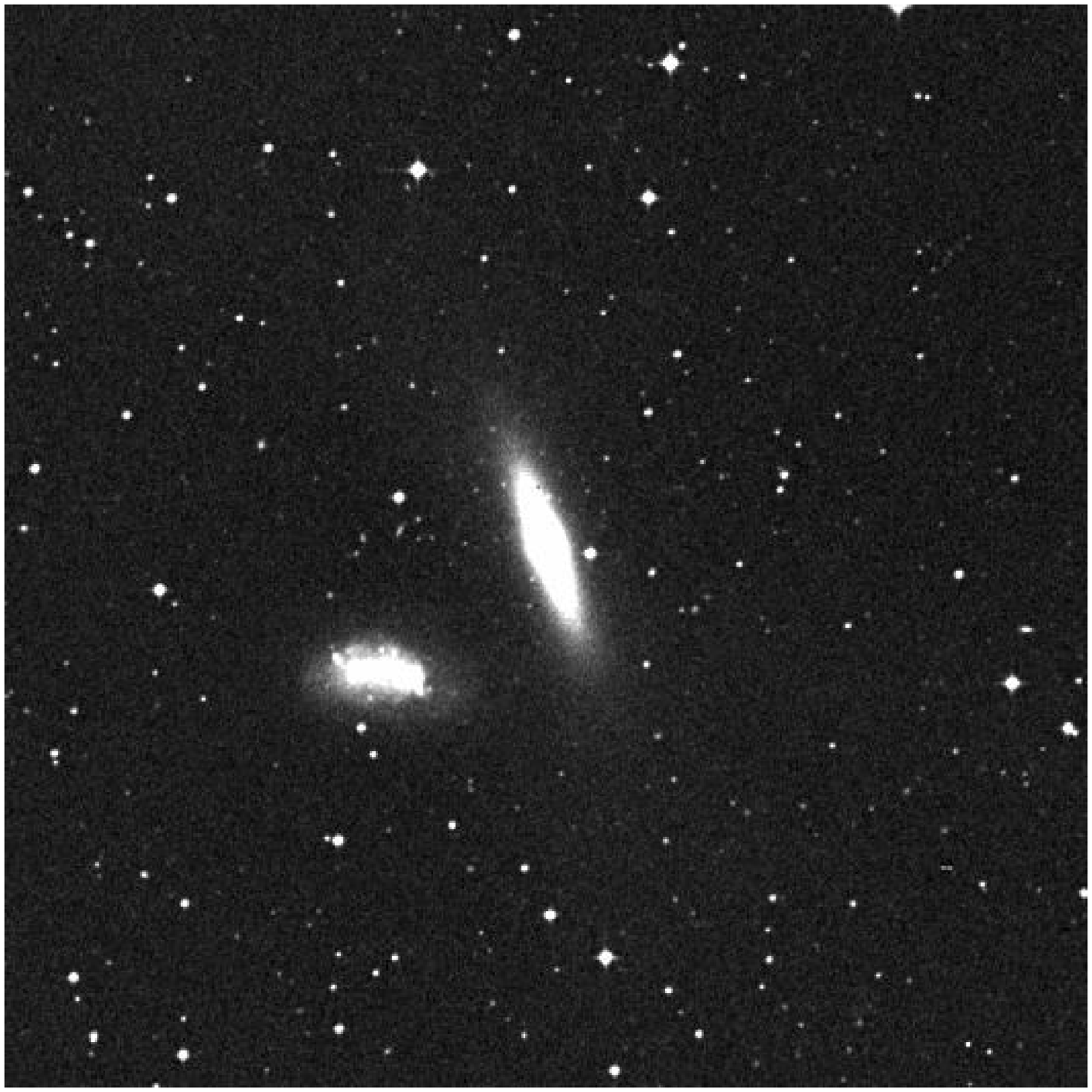}\\
\vspace*{0.5cm}
\includegraphics[width=6.9cm]{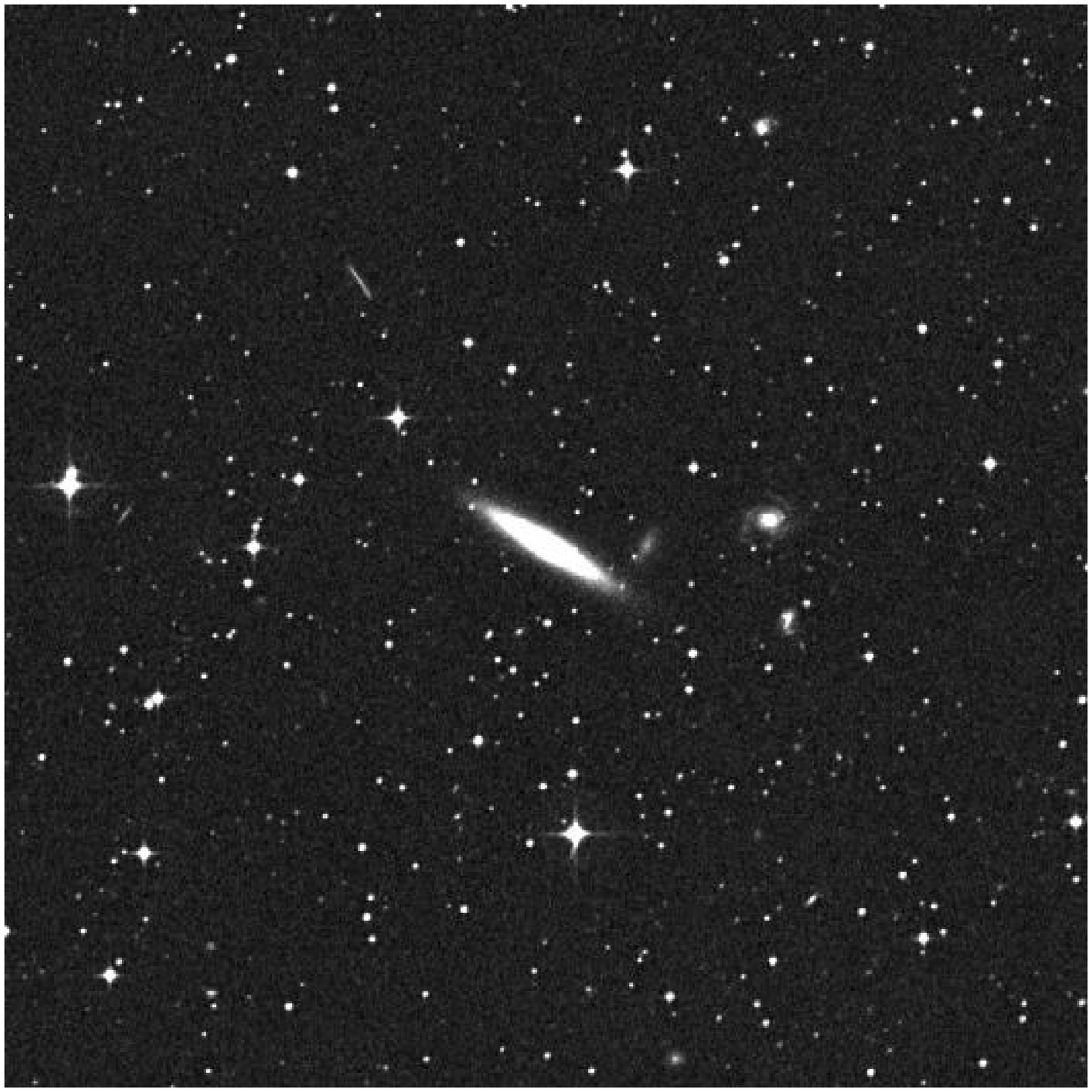}
\caption{Immediate environment of NGC~128 (top), NGC~1596 (middle) and
NGC~3203 (bottom). Each panel shows a $15$~arcmin~$\times$~$15$~arcmin
DSS image centred on the galaxy, with the usual orientation of North
up and East to the left.}
\label{fig:env}
\end{center}
\end{figure}

To gauge the immediate environment of the galaxies,
$15$~arcmin~$\times$~$15$~arcmin DSS images of NGC~128, NGC~1596 and
NGC~3203 are shown in Figure~\ref{fig:env}.
%
% Figure: Environment
%
%
% Galaxies
%
\section{THE GALAXIES}
\label{sec:galaxies}
\subsection{NGC~128}
\label{sec:ngc128}
NGC~128 (S0 pec) is the prototype of galaxies with a B/PS bulge, being
recognized as such already in \cite{bb59}, \cite{s61}, and
\cite{hm66}. Good quality optical and near-infrared (NIR) surface
photometry is available in \cite{cdc91} and \cite{tdfdw94},
respectively, as well as later publications
\citep[e.g.][]{ea97,ocmzb99}. \cite*{sdb90} attempted to quantify the
main global parameters describing the peanut shape of the bulge
\citep[but see also][]{sd00a,sd00b,sd01,sga04}.

\cite{bc77} first discussed the stellar rotation curve of NGC~128 and
detected its ``double-hump'' structure, nicely confirmed by
\cite{ocmzb99}. Although ionized-gas emission had already been
detected, \cite{kfm96} first pointed out that it is counter-rotating.
\cite{ea97} and \cite{ocmzb99} published comprehensive studies of the
structure and stellar/ionized-gas kinematics of NGC~128, confirming
the counter-rotation of the ionized gas and showing that it is tilted
by $\approx25$--$40\deg$ with respect to the main (equatorial) plane
of the galaxy \citep[see also][]{mefpt95,opcm97}. The main disc and
B/PS structure have largely identical and uniform colours \citep[see
also][]{tdfdw94}, but the counter-rotating disc is visible as a slight
reddening in the central regions and is thus probably dusty. Relying
on N-body simulations \citep[e.g.][]{cs81,cdfp90}, the peanutness of
the isophotes, the counter-rotating but tilted ionized-gas disc (see
Section~\ref{sec:bars}) and the gaseous cylindrical rotation,
\cite{ea97} argued that the peculiar shape of NGC~128 is due to an
edge-on bar viewed nearly side-on. \cite{ocmzb99} also convincingly
showed cylindrical rotation in the stars, as expected for barred
galaxies viewed edge-on \citep[e.g.][]{cdfp90}, and a plateau in the
major-axis light profile, traditionally associated with edge-on bars
(see, e.g., \citealt{ba05} and references therein). \cite{cb04} later
used the characteristic shape of the stellar kinematic profiles ($V$,
$\sigma$, $h_3$ and $h_4$) to argue along similar lines.

NGC~128 is in a compact group and is clearly interacting with its
neighbors NGC~127 (SA0$^0$; $\Delta V\approx190$~km~s$^{-1}$) and
NGC~130 (SA0$^-$; $\Delta V\approx190$~km~s$^{-1}$). \cite{bvc90}
briefly explored the dynamics of the triplet. The galaxies NGC~125
((R)SA0$+$ pec; $\Delta V\approx1065$~km~s$^{-1}$) and NGC~126
(SB0$^0$; $\Delta V\approx10$~km~s$^{-1}$), slightly farther away but
also visible in Figure~\ref{fig:env}, are also part of the same group
(although NGC~125's redshift is rather high). To our knowledge, no gas
is detected within the central parts of the group except for NGC~128's
counter-rotating disc. Dust is however present and peaks where NGC~128
interacts with NGC~127 ($\ga6\times10^6$~M$_{\sun}$ of dust;
\citealt{ocmzb99}) and NGC~125 has an H$\,${\small I} detection
\citep{cbf87}. New improved H$\,${\small I} observations will also be
presented in a future publication (in preparation).

A search within a projected radius $R=0.5$~Mpc from NGC~128 using NED
reveals many other possible companions, most with unknown redshifts
but a number with similar ones: UGC~286 (Sbc; $\Delta
V\approx230$~km~s$^{-1}$; $R=0.19$~Mpc), IC~17 (S; $\Delta
V\approx75$~km~s$^{-1}$; $R=0.30$~Mpc), UGC~281 (Scd; $\Delta
V\approx55$~km~s$^{-1}$; $R=0.42$~Mpc), LSBC~F824-10 (dI; $\Delta
V\approx175$~km~s$^{-1}$; $R=0.45$~Mpc), UGC~277 (Scd; $\Delta
V\approx120$~km~s$^{-1}$; $R=0.47$~Mpc) and UGC~275 (SAB(rs)b pec;
$\Delta V\approx145$~km~s$^{-1}$; $R=0.50$~Mpc). Many are H$\,${\small
I}-rich. See also \cite{ga93}, \cite{gmcp00}, \cite{fh02} and
\cite{rgpc02} for studies of group membership.
\subsection{NGC~1596}
\label{sec:ngc1596}
Little detailed information on NGC~1596 is available in the
literature. As for NGC~128, \cite{sdb90} quantify the global
parameters describing the boxy shape of the bulge, while \cite*{jfk95}
present more general global results, also from optical surface
photometry. A more detailed analysis by \cite{pbld04}, who attempted
but failed to fit a thin plus thick disc model, reveals an extended,
roughly speroidal but somewhat distorted outer envelope, which they
suggest might be related to an interaction with NGC~1602 (see
below). \cite{bg90} obtained both optical and NIR photometry and
showed that, like most of their S0 sample, the NIR-NIR colors of the
bulge and disc of NGC~1596 are similar, indicating a similar
metalicity, while the optical-NIR colors are significantly different,
suggesting a younger disc (in the mean).

The only spatially resolved kinematic observations available for
NGC~1596 are the current ionized-gas kinematics and the stellar
kinematics of \cite{cb04}, who argued that the B/PS bulge
of NGC~1596 is due to an edge-on bar viewed nearly end-on.

NGC~1596 forms a pair and is most likely interacting with the
H$\,${\small I}-rich galaxy NGC~1602 (IB(s)m pec; $\Delta
V\approx60$~km~s$^{-1}$) only $3.0$~arcmin away and easily visible in
Figure~\ref{fig:env}. There are again many other possible companions
within a $0.5$~Mpc projected radius, but only three galaxies have a
similar known redshift: NGC~1581 (S0$^-$; $\Delta
V\approx90$~km~s$^{-1}$; $R=0.13$~Mpc), ESO~157-G~030 (E4; $\Delta
V\approx40$~km~s$^{-1}$; $R=0.25$~Mpc), and NGC~1566
(R$^\prime$SAB(rs)bc; $\Delta V\approx5$~km~s$^{-1}$;
$R=0.34$~Mpc). See also \cite{mcl89}, \cite{gmcp00} and \cite{kkfbm05}
for studies of group membership. NGC~1596 was not detected by {\it
IRAS} but is apparently H$\,${\small I}-rich
\citep[e.g.][]{rmgws82,kkfbm05}, although it is unlikely that a
single-dish telescope could disentangle its flux from that of
NGC~1602. Recently acquired H$\,${\small I} synthesis observations
will clarify the issue \citep{ckbg06}.
\subsection{NGC~3203}
\label{sec:ngc3203}
No in-depth study of NGC~3203 exists in the literature, although
\cite{gp97} present broadband optical imaging and argue for an
increasing disc scaleheight with (projected) radius, which they
attribute to the gradual dominance of a thick disc over the thin
disc. \cite{gp00} however show that NGC~3203 has no discernable
vertical colour gradient. \citeauthor*{gpk97}'s (\citeyear{gpk97})
$K^\prime$-band data also suggest a change in the steepness of the
vertical light distribution near the centre, contrary to most galaxies
in their sample. Because of the particular shape of the $K^\prime$
light profiles along cuts parallel to but offset from the major-axis,
which display a characteristic plateau and secondary maxima,
\cite{ldp00} argue that NGC~3203 is barred.

While little is known about the dynamics of NGC~3203, except for the
ionized-gas kinematics presented here and the stellar kinematics of
\cite{cb04}, the latter also argue that the B/PS bulge of NGC~3203 is
due to an edge-on bar viewed nearly end-on.

NGC~3203 has many possible companions within $R=0.5$~Mpc, but none
with a known similar redshift: NGC~3208 (SAB(rs)bc; $\Delta
V\approx470$~km~s$^{-1}$; $R=0.5$~Mpc) has a rather high systemic
velocity. NGC~3203 was not detected by {\it IRAS} or in H$\,${\small
I}, but new improved H$\,${\small I} observations will appear in a
future publication (in preparation).
%
% Discussion
%
\section{DISCUSSION}
\label{sec:discussion}
\subsection{Incidence of gas counter-rotation}
\label{sec:incidence}
\cite{bbz92} presented the first rigorous statistical study of
counter-rotating or strongly kinematically decoupled ionized gas in
disc galaxies \citep[see also][]{bccrz95}. They identified $3$ gas
counter-rotators in a sample of $15$ bright and nearby S0s with
extended ionized-gas ($20\pm10\%$ for $1$ standard deviation assuming
a binomial distribution), suggesting that at least $40\%$ of such
objects may have acquired their gas externally (assuming randomly
oriented infall). Given that pre-existing co-rotating gas (e.g.\ due
to stellar mass loss) will tend to decrease the fraction of observed
counter-rotating systems (due to shocks and the associated loss of
angular momentum), the true fraction of objects with significant
external gas accretion is probably even higher. Although it does not
change their conclusions, it is worth noting that \cite{bbz92} missed
the gas counter-rotation in NGC~128 and that the gas distribution and
kinematics in NGC~2768 is more nearly perpendicular to the major-axis
\citep{fi94}.

\cite{kfm96} later studied a sample of $28$ highly inclined S0s
galaxies in a variety of environment, searching for counter-rotating
stars. While no new system with counter-rotating stars was found, $4$
of the objects displayed counter-rotating ionized-gas. This is
$14\pm7\%$ of the entire sample or $24\pm10\%$ of the objects with
detected emission.

\cite{kf01} also searched for bulk counter-rotating ionized-gas in a
sample of $67$ galaxies from all morphological types, extracted from
the Nearby Field Galaxy Survey \citep{jffc00} to have both stellar and
ionized-gas rotation curves available \citep{k01}. Those galaxies are
generally fainter than in the above studies but better represent the
local galaxy population. A total of $4$ E/S0 gas counter-rotators were
found, representing $7\pm3\%$ of the entire early-type sample or
$24\pm10\%$ of those with both stellar and ionized-gas rotation
curves. If only galaxies currently classified as S0s in NED are
considered (including E/S0 and S0/a galaxies), the fractions become
$6\pm4\%$ and $18\pm12\%$, respectively. Only one example of peculiar
kinematics was found amongst $38$ Sa-Sbc spirals, although in all
cases the authors correctly argue that the number of counter-rotating
objects identified represents a rather stringent lower limit.

\cite{pcvb04} searched for counter-rotation in a sample of $50$
relatively bright and nearby S0/a--Scd galaxies for which major-axis
stellar and ionized-gas kinematics are available from the literature
(using similar observations and data analysis methods; see
\citealt{bccpps96,cetal99,vpcfzbb01,cpcb03}). They detect $2$ gas
counter-rotators only (they ignore the apparent counter-rotation in
the outer disc of NGC~7213; see \citealt{cpcb03}), all of which are
S0s according to NED, for a fraction of $4\pm3\%$. If only galaxies
currently classified as S0s are considered, the fraction becomes
$20\pm13\%$. The low fraction of counter-rotators in intermediate and
late-type spirals in this and the above study probably simply reflects
again the fact that, in those systems, the counter-rotating gas is
swept away by the dominant co-rotating one (independently of its
origin).

In the current work, we detect $3$ galaxies with counter-rotating
ionized-gas in a sample of $30$ objects. This represents $10\pm5\%$ of
the entire sample, but the objects cover the morphological types
S0--Sbc rather inhomogeneously, as they were primarily selected to
cover a wide range of B/PS bulge morphologies. If we restrict
ourselves to the objets classified as S0s in NED, the sample decreases
to $14$ but the $3$ counter-rotating objects remain. They thus
represent a fraction of $21\pm11\%$ of all the S0s, or $27\pm13\%$ of
those where ionized-gas was detected. Those fractions are slightly
higher than the aforementioned studies, presumably because of the
better quality of the data, but they are consistent within the errors.

Merging the $5$ samples discussed above, and keeping only galaxies
currently classified as S0s in NED, we obtain a sample of $94$
objects. Correcting obvious kinematic mistakes from the recent
literature, but {\em not} carrying out a full literature search for
each object, we find that $34$ of those do not have detectable
ionized-gas. Of the $60$ objects that do, it is counter-rotating in
$14$. The fraction of objects with counter-rotating ionized-gas is
thus $15\pm4\%$ for the entire S0 sample, or $23\pm5\%$ for the
objects with detected ionized-gas only. Those fractions are again
consistent with previous studies, but the errors have shrunk
significantly.
\subsection{Gas infall versus minor mergers}
\label{sec:infall_mergers}
External gas acquired by a disc galaxy will generally settle in the
equatorial plane, but it can also settle in the meridional one,
forming a polar-ring or polar-disc structure
\citep[e.g.][]{td82,sd82,ckrh92}. Once equilibrium has been reached,
those configurations lead to co- or counter-rotating and perpendicular
kinematics, respectively. Although the (normalised) cold and warm gas
content of polar-rings is much higher than that of counter-rotators
(the latter being similar to normal galaxies), this probably reflect
different dynamical evolutions rather than different origins (i.e.\
massive self-gravitating polar-rings can be stable;
\citealt{s86,as94,bggr01a}).

\cite*{bbg97} studied the environment of polar-ring galaxies and found
it to be normal, in the sense of the closeness, size, and likelihood
of interaction with nearby objects. Interestingly, there is at least
one similarly-sized companion near almost all objects. An analogous
study of galaxies with counter-rotating gas or stars by \cite*{bgp01b}
(who also considered the local density of galaxies on different
spatial scales) finds similar results. While, superficially, this
seems to indicate that counter-rotating material does not originate
from recent interactions, it may also simply indicate that
counter-rotation develops in rather standard conditions, and that the
specific characteristics leading to counter-rotation (e.g.\ specific
orbital configurations) are not captured by the coarse parameters
considered.

From the ionized-gas statistics discussed above, external gas
accretion is clearly a widespread phenomenom in S0 galaxies. A similar
conclusion can be reached from an analysis of their neutral hydrogen
(H$\,${\small\rm I}) and dust content \citep[e.g.][]{wk86,f91}. The
question remains, however, whether the external gas is accreted slowly
and continuously (gas infall) or sporadically in lumps (minor
interactions and mergers).

Using numerical simulations, \cite{tr96} argue that gas infall is best
suited for the creation of {\em massive} counter-rotating discs in
spirals. Indeed, as long as it is not too clumpy and its infall rate
is not too high, infalling gas will not unreasonably disturb and
thicken the pre-existing galactic disc, while its angular momentum
will determine the scalelength and formation time-scale of the
counter-rotating disc. Furthermore, the clumpier the infalling gas,
the smaller the counter-rotating disc (due to dynamical friction).

For a successful merging scenario, the satellite galaxy should be
significantly less massive and dense than the host, otherwise a more
spheroidal system will emerge and/or the disc will significantly
thicken \citep*[e.g.][]{hb91,b92,qhf93,b98,bg98}. While this is
ill-suited for the formation of {\em massive} counter-rotating discs
(as the merging time-scale is long and many such events are required
to build up the mass; see, again, \citealt{tr96}), it is probably
befitting for the formation of the majority of counter-rotating discs
which are spatially limited and thus (presumably) rather light,
especially if the merging satellites are gas dominated.

We do not separate here between events where the satellite galaxies
actually merge and events where simple tidally-induced transfers of
material take place, since those are of a similar nature. It is
nevertheless clear that tidally-induced gas accretion without merging
is easier to reconcile with the requirement of keeping the host galaxy
disc cold. Based on numerical simulations, \cite{bc03} also argue in
favor of accretion rather than merging in the case of polar-rings.

The presence of (likely) companions near all three galaxies discussed
here, and the apparent disturbed structure of the counter-rotating gas
in at least two of them, suggest that the counter-rotating gas was
indeed accreted in one or perhaps at most a few discrete events. Other
authors argue along similar lines for different objects (e.g.\
\citealt{hjbbm00} for Sa galaxies). Based on the paucity of plausible
nearby companions for their counter-rotators, \cite{kf01} argue
against tidally-induced mass transfer and favor (single) mergers, but
they recognize that possible culprits may be hard to identify. In
fact, the higher H$_2$/H$\,${\small\rm I} ratio observed in
counter-rotating spirals (compared to polar-rings and normal galaxies)
argues for the accretion of external material, as there should be
increased atomic to molecular gas conversion near the co- to
counter-rotating transition region \citep{bggr01a}.

H$\,${\small\rm I} synthesis observations may actually represent the
best hope of identifying ``smoking gun'' evidence for external gas
accretion, since both faint galaxies and tidal features are most
easily detected in H$\,${\small\rm I}. Data for a significant number
of ionized-gas counter-rotators may thus go a long way toward
identifying the origin of the counter-rotating material (or at least
ruling out some possibilities). H$\,${\small\rm I} synthesis
observations of the current sample of $3$ counter-rotators have
already been acquired and will be discussed in future publications
(e.g.\ \citealt{ckbg06} for NGC~1596).
\subsection{Counter-rotating gas and bars}
\label{sec:bars}
Given that the counter-rotating gas in NGC~128, NGC~1596 and NGC~3203
was discovered during a study of B/PS bulges, in which \cite{cb04}
argued that the B/PS bulges of all three galaxies are in fact thick
edge-on bars, it is worthwhile to ask if the counter-rotation itself
might help us understand the structure, formation and evolution of
those bulges.

As argued above, it is likely that the counter-rotating gas in all
three galaxies was accreted from nearby objects during tidal
interactions. It is thus entirely possible that the bars argued to be
at the origin of the B/PS bulges were triggered by those same
interactions. Indeed, it has been firmly established through numerical
simulations that bar formation can be triggered or accelerated by
gravitational interactions \citep*[e.g.][]{n87,gca90,mn98}. Standard
bar-driven evolution (e.g.\ gas inflows, buckling, redistribution of
angular momentum) may then take over, most likely driving the galaxies
toward earlier types and the bulges toward boxy and peanut
shapes. \cite{mwhdb95} in fact present a simulation where a galaxy
satellite is acccreted, triggering the formation of a bar which
subsequently evolves into a B/PS bulge. Such a scenario is thus
possibly at play in NGC~128, NGC~1596, and NGC~3203. The same
interactions might also trigger bar formation in the satellite
galaxies, if they are not accreted or destroyed in the process. This
may in fact be the case in NGC~1596's companion NGC~1602, which is
also barred. Independently of bar-driven evolution, it is of course
also possible that minor mergers will lead to bulge growth and disc
thickening, again driving the host towards earlier types.

The counter-rotating gas discs may also provide an independant test as
to whether NGC~128, NGC~1596, and NGC~3203 are truly barred. Indeed,
only rotating triaxial potentials possess stable counter-rotating
periodic orbits which are inclined with respect to the equatorial
plane (e.g., the so-called anomalous orbits;
\citealt*{m82,hms82,mh84}). If the counter-rotating gas in the three
galaxies is settled, as seems to be the case in NGC~128 and possibly
in NGC~1596, then it must lie in such a configuration \citep[see
also][]{fu93}. Using integral-field spectroscopy, \cite{ea97} clearly
showed that this is the case in NGC~128, thus supporting the barred
origin of its B/PS bulge, but NGC~1596 and NGC~3203 lack both
integral-field spectroscopy and narrow-band imaging.

Although the above argument has not received much attention in the
literature, integral-field and Fabry-Perot observations clearly show
that the counter-rotating ionized-gas discs in the (nearly) edge-on S0
galaxies NGC~4546 \citep{pbacm98,setal05} and NGC~7332
\citep{pb96,fetal04} are also inclined, although they are also rather
disturbed. Both are believed to be barred
\citep[e.g.][]{sb94,fetal04}.

We also note that \cite{f96} and \cite{dh97} discuss the development,
evolution and kinematic signatures of counter-rotating bars in the
presence of counter-rotating stars \citep[see also][]{sm94}.
%
% Conclusions
%
\section{CONCLUSIONS}
\label{sec:conclusions}
Using our previous studies of the ionized-gas and stellar kinematics
of a relatively large sample of $30$ edge-on disc galaxies with
(mostly) boxy and peanut-shaped (B/PS) bulges, we have identified two
new galaxies (NGC~1596 and NGC~3203) where the ionized gas is
counter-rotating with respect to the bulk of the stars. We have also
confirmed similar kinematics in one additional object
(NGC~128). Counter-rotating gas is thus present in $10\pm5\%$ of our
entire sample. However, if only lenticular (S0) galaxies are
considered, the fraction climbs to $21\pm11\%$. This fraction further
climbs to $27\pm13\%$ if only S0s with extended ionized-gas are
considered. As discussed at length in the text, those fractions are
consistent with but slightly higher than the few existing systematic
studies available in the literature. Merging those studies, fractions
of $15\pm4\%$ and $23\pm5\%$ are obtained for, respectively, all S0s
and S0s with ionized-gas only.

Based on the presence of probable companions near the galaxies
discussed here, we argued that minor mergers and tidally-induced
transfer of material from nearby objects are primarily responsible for
the counter-rotating gas. If accretion on to the objects is isotropic,
similar processes must have been at work in roughly twice the
fractions of objects discussed. This strongly argues for a
non-negligible role of (minor) accretion in galaxy formation and
evolution.

The presence of counter-rotating gas in barred galaxies offers a
number of largely unexplored ways to probe the structure and dynamics
of those objects. In edge-on galaxies in particular, where
counter-rotation is easiest to detect, it offers an independent way of
testing the barred nature of B/PS bulges.
%
% Acknowledgments
%
\section*{Acknowledgments}
We wish to thank J.\ van Gorkom, B.\ Koribalski, E.\ Athanassoula, G.\
Aronica, A.\ Bosma, and K.\ C.\ Freeman for useful discussions at
various stages of this work. MB acknowledges support by NASA through
Hubble Fellowship grant HST-HF-01136.01 awarded by Space Telescope
Science Institute, which is operated by the Association of
Universities for Research in Astronomy, Inc., for NASA, under contract
NAS~5-26555, during part of this work. AC was supported by NSF grant
AST~00-98249 to Columbia University. The Digitized Sky Surveys were
produced at the Space Telescope Science Institute under U.S.\
Government grant NAG~W-2166. The images of these surveys are based on
photographic data obtained using the Oschin Schmidt Telescope on
Palomar Mountain and the UK Schmidt Telescope. The plates were
processed into the present compressed digital form with the permission
of these institutions. This research also made use of NASA's
Astrophysics Data System (ADS) Bibliographic Services, of the
NASA/IPAC Extragalactic Database (NED), which is operated by the Jet
Propulsion Laboratory, California Institute of Technology, under
contract with the National Aeronautics and Space Administration, and
of HyperLEDA.
%
% Bibliography
%

%
\end{document}